\nonstopmode
\documentstyle[epsf]{article}
\title{Imaging algorithms in radio interferometry}
\author{R.J. Sault \and T.A. Oosterloo}
\date{}
\begin{document}
\maketitle

\newcommand{\arcsec}{\prime\prime}
\newcommand{\arcmin}{\prime}
\newcommand{\degr}{^{\circ}}

\section{Introduction}
Unlike an optical telescope, the basic measurements of a radio
interferometer (correlations between antennas) are indirectly
related to a sky brightness image. In a real
sense, algorithms and computers are the lenses of a radio interferometer.
In the last 20 years, whereas
interferometer hardware advances have resulted in improvements of a factor of
a few, algorithm and computer advances
have resulted in orders of magnitude improvement in image
quality. Developing these algorithms has been a fruitful
and comparatively inexpensive method of improving the performance of existing
telescopes, and has made some newer telescopes possible.
In this paper, we review recent developments in the algorithms
used in the imaging part of the reduction process.

What constitutes an `imaging algorithm'? Whereas once there was a steady
`forward' progression in the reduction process of editing, calibrating,
transforming and, finally,
deconvolving, this is no longer true.
The introduction of techniques such as self-calibration, and algorithms that go
directly from visibilities to final images, have made the dividing lines
less clear. Although we briefly
consider self-calibration, for the purposes of this paper calibration issues are generally excluded.
Most attention will be directed to the
steps which form final images from the calibrated visibilities.

\section{The van Cittert-Zernike equation}
To set the framework, we briefly summarize the fundamental theory of radio
interferometric imaging. See {\it Thompson et al.} [{\it 1986}] or {\it Perley et al.} [{\it 1989}]
for comprehensive descriptions.

Assuming a quasi-monochromatic, spatially incoherent source, the response of an
interferometer is given by the van Cittert-Zernike equation:
\begin{equation}
V(u,v,w) =
 \int \frac{A(\ell,m)I(\ell,m)}{n}\exp\left(-i2\pi(\ell u+mv+(n-1)w)\right)d\ell\,dm.
\end{equation}
Here $(u,v,w)$ is the spacing (the distance between antenna pairs measured in
wavelengths)
and $(\ell,m,n)$ are direction cosines of the brightness distribution with respect
to this coordinate frame (note that $n=\sqrt{1-\ell^2-m^2}$). The coordinate
system is defined such that the $w$-axis points in the direction
of the source.

The van Cittert-Zernike equation indicates that the complex-valued visibility
function, $V(u,v,w)$, is a Fourier-like integral of the sky brightness,
$I(\ell,m)$, multiplied by the primary beam response of an interferometer,
$A(\ell,m)$, and $1/n$. Most radio interferometric telescopes have more
than two antennas, and so multiple spacings
can be made simultaneously. Indeed for $N$ antennas, $N(N-1)/2$ spacings
can be measured. Additionally, as the coordinate system is
one fixed on the sky, the Earth's spin causes the measured spacings to
rotate during a synthesis. Thus a good coverage in the $u-v-w$
domain can be achieved. Further coverage can be gained by physically
moving the antennas.

For a number of reasons (mainly computational simplicity) it has been
conventional to make a small field assumption, i.e. $n\approx1$.
In this case, ignoring the primary beam term, the van Cittert-Zernike
equation becomes independent of $n$ and $w$, and reduces to a two-dimensional
Fourier transform:
\begin{equation}
V(u,v) = \int I(\ell,m)\exp\left(-i2\pi(\ell u+mv)\right)d\ell\,dm.
\label{eq:2D}
\end{equation}
The primary beam term can be ignored -- its effect can be subsumed into the
estimate of the sky brightness, and can be corrected for during
the analysis stage of the final image.
Historically, this two-dimensional relationship has been adequate (or
tolerated!) for nearly all
radio interferometric imaging. Using this relation 
for imaging is the conventional
approach. However, much recent research has been devoted to devising
ways of avoiding the assumptions and approximations in this formulation, and
thus avoiding errors or allowing larger fields to be imaged.

In this context, this paper can be roughly divided into two parts. The first
part briefly reviews conventional imaging and discusses some
recent refinements. The second part
addresses recent research to extend imaging beyond the
small-field approximation.
The two parts are far from
mutually exclusive. The extensions are invariably based on the 
conventional approach in some way.
Consequently, refinements in the conventional approach flow
immediately through to the extensions. 

\section{Conventional imaging}
Whereas
Equation (2) 
suggests that an inverse two-dimensional
Fourier transform will recover a sky brightness,
in practice the visibility function is sampled at only a
discrete set of points, $(u_j,v_j)$. So the inverse transform must be
approximated by
\begin{equation}
I_{\rm D}(\ell,m) = \sum_j w_j V(u_j,v_j) \exp[i2\pi(u_j \ell + v_j m))].
\label{eq:dft}
\end{equation}
Here $w_j$ is a weight assigned to each visibility measurement (discussed
below), and $I_{\rm D}(\ell,m)$ is called the dirty image.
In practice, direct implementation of
Equation (3)
is rarely done -- the
Fourier transform operation is usually implemented with FFTs. As this requires the
visibilities to be on a regular rectangular grid, a gridding step is required.
This is usually done by convolving 
the visibilities with a particular function and resampling the result at the FFT
grid points (other techniques have been used; see {\it Thompson
and Bracewell}, {\it 1974}). The convolving function is chosen to suppress the aliased power
in the image. See {\it Sramek and Schwab} [{\it 1989}] for details.

We define the dirty (or synthesized) beam, $B(\ell,m)$, as the image
that results after
replacing $V(u_i,v_j)$ by 1 in the Fourier transform step. The
dirty beam is the
response of a point source at the phase centre. Provided
Equation (2)
is an adequate approximation, the dirty image will be the convolution of the 
true sky brightness with the dirty beam,
\begin{equation}
I_{\rm D}(\ell,m) = I(\ell,m) \ast B(\ell,m).
\end{equation}
That is, the dirty beam is the shift-invariant point-spread function (PSF)
of the imaging system.
In later Sections, where deviations from
Equation (2)
are considered,
the true PSF will differ from the dirty beam.

The following Sections consider recent research in weighting the visibility
data before the Fourier transform step, and in the deconvolution algorithms used
to estimate the true sky brightness from the dirty images.

\subsection{Weighting}
Weighting by $w_j$ in the Fourier transform operation is required to
account for the different sampling densities in the Fourier plane.
Although there are variants [see {\it Sault}, {\it 1984}; {\it Sramek and Schwab}, {\it 1989}], there
are two traditional approaches:
\begin{itemize}
\item Natural weighting is used to maximize point-source sensitivity in the
dirty image. Here the visibility weight is inversely proportional to the 
noise variance of the visibility.
\item Uniform weighting is used to minimize sidelobe levels. The weights
are inversely proportional to a local density function.
\end{itemize}
In the usual case where there is an excess of shorter spacings, natural
weighting produces significantly poorer dirty beam shape and resolution than
uniform weighting. Conversely the noise variance in a uniformly weighted
dirty image will typically be a factor of a few worse than the
naturally weighted image.

{\it Briggs} [{\it 1995}] has developed a new scheme for weighting visibilities,
called {\sl robust weighting}, which
can offer a significant improvement.
This scheme minimizes the expected rms difference (either as a result of
sidelobes or noise) between the dirty image and the true sky for a point
source. It is closely related to the Wiener minimum mean-squared error
criteria.
For high values of the sampling density function or if the
signal-to-noise ratio is very good, robust weighting reduces to uniform weighting.
Conversely, in sparsely sampled regions of the visibility function or for poor
signal-to-noise ratio, it reduces to natural weighting -- this prevents excessive
noise amplification.

To some extent, robust weighting combines the best of both natural and uniform
weighting. It can reduce the noise in an image, compared to uniform weighting,
without much loss of resolution. For example, for a full-track VLA
observation, it can reduce the noise by 25\%, while the width of the beam
increases by only 3\%. In cases such as unbalanced VLBI arrays, the improvement
in noise can be as significant as a factor of 2, with a loss of resolution of only about
15\%. An example for the ATCA is given in
Figure 1.

\begin{figure}
\begin{center}\epsfysize=8cm\leavevmode\epsffile{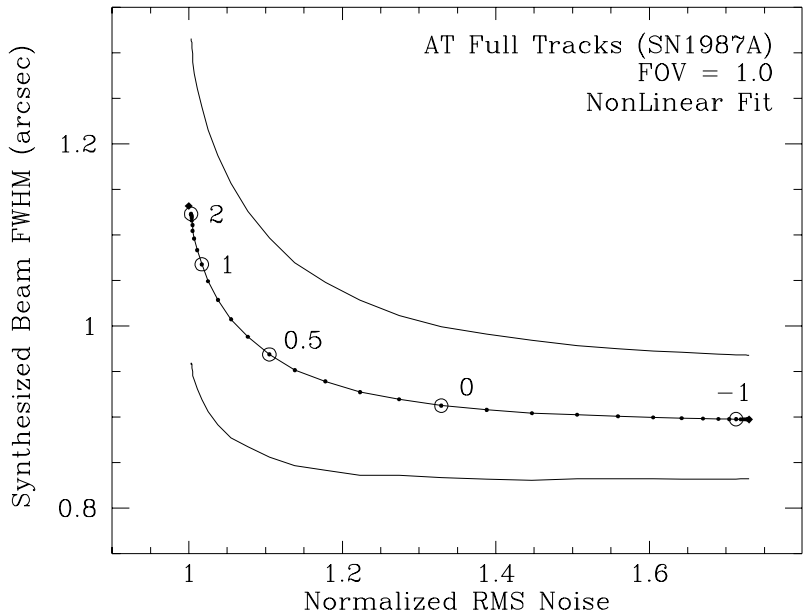}\end{center}
\caption[]{
Thermal noise and resolution as functions of robustness for a full configuration
of the ATCA. The three lines give the major, minor and geometric mean of the
dirty beam FWHM. The dots on the geometric mean (centre) curve indicate the
value of the robustness parameter (roughly the logarithm of a signal-to-noise
ratio). Robustness parameters from -2 to 2 vary the weighting between
approximately uniform and natural weights. Reproduced from {\it Briggs} [{\it 1995}].}
\label{fig:robust}
\end{figure}

\subsection{Deconvolution}\label{sec:deconv}
The most commonly used approaches to deconvolving the dirty image fall into two
groups -- CLEAN algorithms and maximum entropy methods (MEM).

The original CLEAN algorithm was introduced by {\it H\"ogbom} [{\it 1974}; see also
{\it Schwarz}, {\it 1978}, {\it 1979} and {\it Cornwell and Braun}, {\it 1989}].
In the basic CLEAN
algorithm the maximum, in absolute value, in the dirty image is located. 
Assuming that this peak is the result of a point source, a scaled version of
the dirty beam (i.e. the response of a point source) is subtracted from the
dirty image at this location and the amount subtracted is stored in a
component image. For a uniformly weighted image, this procedure is the optimum
approach (in a least-squares sense) to locating a point source.
This process is repeated on the residual image until the
maximum residual is below a certain cut-off level. The scale of the dirty beam
subtracted is equal to the maximum in the image, multiplied by a relaxation
factor (called the {\sl loop gain}). This procedure results in a sky model
consisting of a set of point sources
({\em components}) and a residual image. Because CLEAN's extrapolation 
of spatial frequencies beyond the maximum spacing is often unreliable,
the image that astronomers
usually analyse goes through a `restore' step.
This consists of summing the residuals and convolving the component
image with a Gaussian (the Gaussian FWHM is chosen to match the dirty beam
FWHM).

A number of variants on the basic CLEAN algorithm have been suggested, either to improve
its efficiency or to alleviate some of its shortcomings.
For example, CLEAN spends much time needlessly checking
small residuals and shifting and scaling the dirty beam. To improve
efficiency, {\it Clark's} [{\it 1980}] variant considers only a list of the largest
residuals as candidate positions and also uses only a small portion of the
dirty beam (the beam-patch) to subtract components. After a certain number of
iterations, the errors in the residuals introduced by this approximation are
eliminated by doing a proper convolution of the current component image with the
dirty beam (implemented with Fourier transforms), and subtracting the result
from the dirty image. W. D. Cotton and F. R. Schwab (reported in {\it Schwab} [{\it 1984}])
take this one step
further. They re-evaluate the residuals by subtracting the components directly
from the visibility data and then re-transforming the residual visibilities.
In this way aliasing because of the convolutional gridding is almost eliminated.
It also allows deconvolving multiple sub-fields in the primary beam
simultaneously (i.e. only the relevant sub-fields need to be imaged if the
primary beam is sparsely filled with emission). 

The second deconvolution approach is based on MEM.
Since only a limited region of the $u-v$-plane is sampled,
an infinite number of images are consistent with the measurements. The maximum
entropy principle suggests that the `best' of these images to select is that which
maximizes the entropy while being consistent with the measured data. Several
forms for entropy have been suggested, although most of the proposed forms give
similar results for most practical applications. The original form
suggested in radio astronomy [{\it Ables}, {\it 1974}; {\it Gull and Daniell}, {\it 1978}] is
\begin{equation}
-\sum_{\ell,m} I(\ell,m)\log( I(\ell,m) ).
\end{equation}
See {\it Cornwell and Braun} [{\it 1989}] and {\it Narayan and Nityananda} [{\it 1986}] for more details. 

To some extent, CLEAN and MEM are complementary in their application. CLEAN
performs well on small, compact sources, while MEM does better on extended
sources. Partly because CLEAN is much easier to understand intuitively and
is more robust, it is more widely used by radio astronomers than MEM.

However, CLEAN does not perform well on very extended sources. Moreover, if
the dirty beam has strong sidelobes, CLEAN can introduce corregations
in the deconvolved
image [{\it Cornwell}, {\it 1983}; {\it Schwarz}, {\it 1984}; {\it Steer et al.}, {\it 1984}]. These appear because
artificial peaks are created by the sidelobes of the dirty beam subtracted at a
nearby position. Severe problems can also be caused by missing short baseline
information. A very extended source can have most of its flux at baselines
shorter than the one measured. The only correct way to handle this is to
measure this short- and zero-spacing information
(see Section 5),
but this is not always feasible and one has to try to
solve the problem in the deconvolution process. The missing information
causes the zero level in the image to vary with position (commonly referred to
as the {\sl negative bowl}). CLEAN is not able to make acceptable images from
these kinds of data. 

Smoothing the dirty image to lower resolution partly helps to overcome some of
these problems, but at the expense of loss of information. 
Multi-resolution approaches have been suggested [{\it Wakker and Schwarz}, {\it 1988}; {\it Starck et al.}, {\it 1994}; {\it Brinks and Shane}, {\it 1984}] that take advantage of smoothing, but
retain the fine-scale structure in the image. {\it Wakker and Schwarz} [{\it 1988}]
have developed the so-called Multi-Resolution CLEAN (MRC) algorithm. Here the dirty
image is smoothed and the difference between the original and the
smoothed dirty image is made. Both the smoothed and the difference image
are deconvolved separately (with the appropriate smoothed- and difference dirty
beams), which ensures that the fine-scale structure is retained. 

A similar approach is given by {\it Starck et al.} [{\it 1994}]: dirty images of different
resolution ranges are made using a wavelet transform and these images are
deconvolved and restored in a way similar to MRC. The wavelet formalism
provides a more rigorous framework for the multi-resolution approach. 

Another approach to the deconvolution problem is to formulate it as a linear
system of the form
\begin{equation}
{\bf Ax} = {\bf b}
\end{equation}
and then use algebraic techniques to solve this system. The elements of ${\bf
A}$ contain samples of the dirty beam, the elements of ${\bf b}$ are samples of
the dirty image, while ${\bf x}$ contains the components of the reconstructed
image. Without any additional constraints, the matrix ${\bf A}$ is singular;
additional information has to be added to find a solution. Assumptions that
regularize the system include positivity and compact support of the source. An
algebraic approach is not new in itself [e.g. {\it Andrews and Hunt}, {\it 1977}], but
practical applications of such techniques on problems of the size met
in radio astronomy have become feasible only recently. 

One approach for solving the linear system is by using a technique called
Non-Negative Least Squares (NNLS). NNLS belongs to the class of least-squares
problems with linear inequality constraints [{\it Lawson and Hanson}, {\it 1974}]. In this
case, the inequality constraints enforce positivity of the source brightness.
Applying
NNLS to deconvolution in radio astronomy has been studied in detail
by {\it Briggs} [{\it 1995}]. One feature of NNLS is that it finds a direct solution,
in contrast to iterative techniques such as CLEAN or MEM.

Although NNLS has its limitations, it performs very well for a certain range of
problems: it does well on compact sources, sources that are too extended for
CLEAN and too small for MEM to be handled properly. NNLS reproduces the
sources with high fidelity, so it works very well as part of a
self-calibration loop.
This is related to the fact that NNLS zeros the residuals in the dirty image
nearly completely. NNLS does have the tendency, like CLEAN, to compact flux.
As with CLEAN, a `restore' step is advisable before analysis.
Interestingly, NNLS can deconvolve an
image from the sidelobes alone, as long as it knows where the source is
supposed to be. 

{\it Lannes et al.} [{\it 1994}] present another technique based on a least-squares
approach where support information is used to regularize the
algorithm. Again a linear system similar to that mentioned above is solved,
but using a technique which iterates between the $u-v$ and image plane.
Unlike CLEAN and NNLS, a `restore' step is not required. Their
technique suppresses excessive extrapolation of higher spatial
frequencies during the deconvolution.

\subsection{Self-Calibration}
Self-calibration has proved a very powerful technique in improving the
ultimate image quality. It is also intimately coupled to some imaging
algorithms. Thus, despite its being somewhat outside the
umbrella of `imaging', some discussion of self-calibration is warranted.
Comprehensive reviews of self-calibration
are given by {\it Pearson and Readhead} [{\it 1984}] and
{\it Cornwell and Fomalont} [{\it 1989}].

Because of differential changes in atmospheric delays to different
antennas and because of amplitude and
phase drifts in the signal paths to the antennas, the measured
visibility, $V_{ij}'$, will be related to the true
visibility, $V_{ij}$, by a complex-valued gain. Because these
effects are antenna-based, the overall gain factor can be
decomposed into two antenna-based gains, $g_i$ and $g_j$, i.e.
\begin{equation}
V_{ij} = g_ig_j^\ast V_{ij}'.
\end{equation}
These gains need to be estimated and the measured visibility corrected
before the imaging stage.
Although regular observations of a calibrator source are normally
used to determine the antenna gains, temporal and spatial variations
in the atmosphere and instrument mean that this approach is limited.

The self-calibration technique has been very successful in solving
this problem: the source under
study is used as its own calibrator. Although this at first appears
unlikely, its validity rests on a number of principles.
\begin{enumerate}
\item Despite antenna gain errors, the data contain good observables.
The best known of these is closure
phase, which is the sum of the visibility phases around a triangle.
It is trivial to show that this is independent of the antenna gains, i.e.
\begin{equation}
\arg(V_{ij}')+\arg(V_{jk}')+\arg(V_{ki}') = \arg(V_{ij})+\arg(V_{jk})+\arg(V_{ki}),
\end{equation}
is a good observable.
\item Whereas there are only $N$ antenna gains, there are potentially
$N(N-1)/2$
baselines. Hence the number of unknown antenna gains is often significantly
smaller than the number of visibilities.
\item Good a priori information is available on the source, such as
compact support and positivity.
\end{enumerate}
Although {\it Jennison} [{\it 1951}, {\it 1958}] used the closure phase relationship quite
early, it was not until it was rediscovered in the 1970s [e.g.
{\it Readhead et al.}, {\it 1980}; {\it Readhead and Wilkinson}, {\it 1978}]
that its use became
widespread. {\it Schwab} [{\it 1980}] and {\it Cornwell and Wilkinson}
[{\it 1981}] further
developed the technique into what is now known as self-calibration.
See {\it Ekers} [{\it 1983}] for a historical prospective.

The first step in a normal self-calibration is to generate a model of
the sky. This is usually done by deconvolving a dirty image (which, in turn,
was made with the best current estimate of the antenna gains). From this
sky model, model visibilities (i.e. the visibilities that
would have been measured if the sky model were correct) can be
generated. These are then used to find new antenna gains that
minimize the difference between the model and measured visibilities.
For example, for model visibilities $V_{ij}$,
\begin{equation}
\epsilon^2 = \sum_{i,j} |V_{ij} - g_ig_j^\ast V_{ij}'|^2\label{eq:selfcal}
\end{equation}
is minimized. Gain solutions are assumed to remain valid over periods of
seconds to hours, depending on physical conditions.

We consider here two more recent contributions to self-calibration
theory.

\subsubsection{Joint redundancy/self-calibration}
Redundancy calibration [{\it Noordam and de Bruyn}, {\it 1982}], like self-calibration,
determines antenna gains directly from the source observations.
However, redundancy calibration requires
that a certain set of the different baselines actually measure the same
spacing, i.e. some baselines are redundant. Redundancy also minimizes
Equation (9),
except that the summation is over only
the redundant baselines, and the model visibility function is also an
unknown: redundancy calibration is model independent. The cost of
this model independence is that, because several baselines are used to
measure the one spacing, $u-v$ coverage suffers.

Another shortcoming of redundant calibration is that it uses only a subset
of the baselines (the redundant ones) when determining antenna gains.
If some of these baselines are particularly noisy, subject to interference,
or otherwise faulty, the redundant solution can be poor.
{\it Wieringa} [{\it 1991}, {\it 1992}] presented a joint self-calibration/redundancy approach
which minimizes
Equation (9).
For the redundant baselines,
the model visibility is treated as an unknown, whereas for the non-redundant
ones, the model visibility is derived from the sky model. Wieringa concluded
that this gives superior performance. He also presented an interesting
analysis of the
merits of redundancy and self-calibration and concluded that redundancy
is probably useful only for arrays with about 10 to 20 antennas.

\subsubsection{Imaging weak sources with poor phase stability}
Self-calibration is applicable only when the
signal-to-noise ratio in a self-calibration solution interval is greater
than about 5. Attempting to self-calibrate noisier data will result in a
significant noise bias: self-calibration will `phase up' the noise to
resemble the model. {\it Cornwell} [{\it 1987}] noted the relationship between
the optical imaging technique of ``speckle masking''
[e.g. {\it Weigelt and Wirnitzer}, {\it 1983}; {\it Lohmann et al.}, {\it 1983};
{\it Lohmann and Wirnitzer}, {\it 1984}; {\it Weigelt}, {\it 1991}] and
closure phase in radio astronomy.
In the radio interferometric context, speckle masking
integrates the quantity
\begin{equation}
V(u,v)V(u',v')V(-u-u',-v-v').
\end{equation}
These three visibilities form a closure triangle. Their product is independent
of atmospheric phase, and the phase of the product is the closure phase of
the source. Integrating allows a good measure of this
to be determined. This technique has found widespread application
in the optical and infrared speckle interferometry and aperture-synthesis
communities [e.g. {\it Haniff et al.}, {\it 1987}; {\it Nakajima et al.}, {\it 1989}; {\it Gorham et al.}, {\it 1989}].
However, its use at radio wavelengths has been limited because of the
narrow range of the regime of
signal-to-noise ratio in the radio where it is the
appropriate method (it is applicable for signal-to-noise ratios
of about 0.5 to a few). Additionally because of the techniques
computational expense and as
interferometers are usually non-redundant,
it is applicable for simple compact objects only.

\section{Wide bandwidth effects}
The van Cittert-Zernike equation assumes quasi-monochromatic radiation,
or essentially that the observing bandwidth is small. When the bandwidth
is significant, interferometers experience `bandwidth smearing' (a
chromatic aberration). This smearing is readily interpreted in the
Fourier plane. For a given antenna separation, the $u-v$ coordinate
will be proportional to frequency. Hence different frequencies within
the bandwidth correspond to different, radially distributed, $u-v$ 
coordinates. The measured correlation is the
integral of the visibility function over these.
That is,
the correlation measures a radial smeared visibility function.
This radial smearing
in the visibility plane corresponds to a radial smearing in the image
plane, with the smearing being proportional to the fractional
bandwidth, $\Delta\nu/\nu$, and the distance from the delay-tracking
centre, $\sqrt{\ell^2+m^2}$. Thus the effect is unimportant for narrow
bandwidths or near the centre of the field of view.
However, the smearing can be quite significant in wide-field images. 
See {\it Cotton} [{\it 1989}] and
{\it Bridle and Schwab} [{\it 1989}] for general discussions.

{\it Clark} [{\it 1975}] noted that appropriate change of coordinates (from $(\ell,m)$
to a polar coordinate system, and then taking the logarithm of the radius)
converts the smearing to a convolution, and so can be handled by a number of
deconvolution algorithms. Such an approach has a number of problems, and does
not appear to have been tried in practice. Small amounts of bandwidth
smearing can be approximately handled in a Cotton-Schwab-like CLEAN algorithm
[e.g. see {\it Condon et al.}, {\it 1994} for one approach]. Alternatively,
{\it Waldram and McGilchrist} [{\it 1990}] used another CLEAN-based approach. Bandwidth smearing
can be seen as an imaging system with a position-variant PSF (the true PSF is
just the normal dirty beam at the phase centre, but becomes a more smeared
version of this as $\sqrt{\ell^2+m^2}$ increases). Waldram and McGilchrist
CLEAN with the true PSF, which they compute (at least approximately) as a
function of position. 

The best approach is to solve the problem in hardware:
the smearing in the $u-v$ plane can be reduced by breaking
the passband into a number of channels.
Correlations are then formed between corresponding channels. Each channel
correlation is then treated as a distinct visibility with its own $u-v$
coordinate.
This reduces the bandwidth smearing to that proportional to the width of
each channel, not the total bandwidth.

Imaging with visibilities measured at a number of frequencies
has become known as multi-frequency synthesis
(and sometimes bandwidth synthesis).
Although we introduce multi-frequency synthesis as a cure for
bandwidth smearing, it is a far more useful technique. In addition to
imaging multiple channels of a passband, multi-frequency synthesis can
be used when the observing frequency is intentionally varied, possibly
by an appreciable amount. Varying the observing frequency is an
alternative way of measuring different $u-v$ locations to physically
moving antennas. It is an attractive technique for both fixed (or
movable arrays where the number of configurations is limited)
if the sampling in the $u-v$ plane would otherwise be sparse.

Although this principle has been known for some time
[e.g. {\it McCready et al.}, {\it 1947}], it has received little
attention until recently. This was probably a result of a lack of
understanding of the image artifacts created by source brightness
changes with frequency. Although the technique is clearly not
appropriate
for spectral-line observations, even for continuum observations there will be
a spectral variation (non-zero spectral index) of source brightness.
{\it Cornwell} [{\it 1984}] appears to have been the
first to consider the resultant image-plane artifacts caused by
spectral variation, and to propose a
deconvolution approach to remove these artifacts. These ideas were developed
further by {\it Conway et al.} [{\it 1990}], {\it Conway} [{\it 1991}] and
{\it Sault and Wieringa} [{\it 1994}].

These deconvolution approaches recognize that if the spectral index of a
point source is
known then its response (a PSF) can be calculated. This will now depend
on both the $u-v$ coverage and sampling in frequency. Furthermore, this
PSF, $B(\ell,m)$, can be expanded as a Taylor's series in
the spectral index,
\begin{equation}
B(\ell,m) = B_0(\ell,m) + \alpha B_1(\ell,m) + \frac{1}{2}\alpha^2 B_2(\ell,m) \ldots.\label{eq:mfsexp}
\end{equation}
Here $B_0(\ell,m)$, $B_1(\ell,m)$, etc, are images that depend only on the $u-v$ and frequency
coverage of the experiment. $B_0(\ell,m)$ is the conventional dirty
beam, whereas $B_1(\ell,m)$ has been termed the `spectral dirty beam' -- it
is the response that results from the linear spectral slope of the source
brightness. Typically $B_1$
has a peak value of the order of a few times $10^{-3}$ for
frequency spreads of $\pm 10$\%. For single-frequency observations, it (and higher
order $B_j$ images) will be identically 0.

Truncating
Equation (11),
the multi-frequency dirty image can be
approximated as obeying a generalized convolution relationship
\begin{equation}
I_{\rm D} = I\ast B_0 + (\alpha I)\ast B_1.
\end{equation}
{\it Conway et al.} [{\it 1990}] have proposed a CLEAN-like algorithm which
alternately iterates with the $B_0$ and $B_1$ beams, thus generating
an estimate of both $I(\ell,m)$ and $\alpha(\ell,m)I(\ell,m)$.
{\it Sault and Wieringa} [{\it 1994}] refine this with an algorithm which deconvolves with
the two beams simultaneously.
Figure 2
gives an example from the latter paper.

\begin{figure}
\begin{center}\epsfxsize=12cm\leavevmode\epsffile{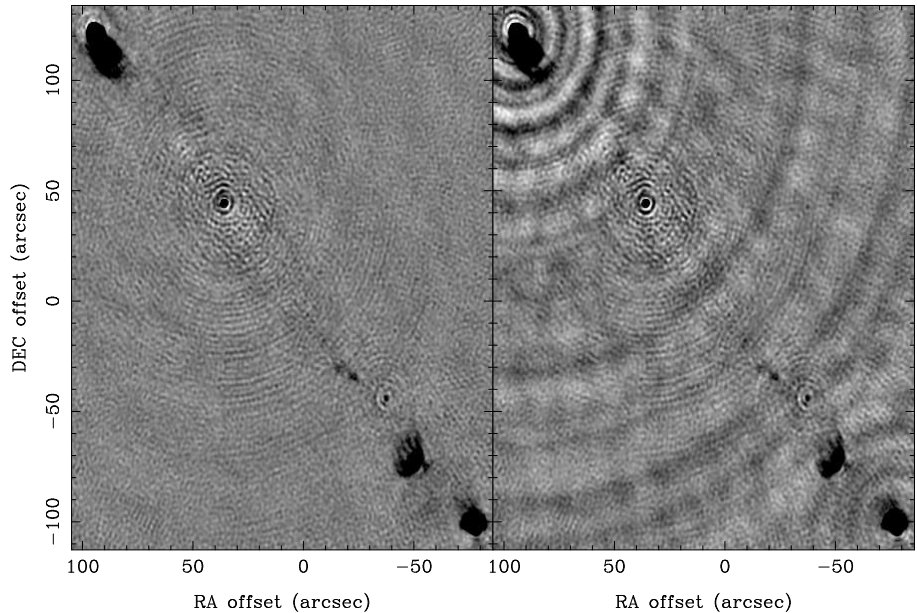}\end{center}
\caption[]{
Two multi-frequency synthesis images made from data with frequencies
spanning 4.4 to 6.1 GHz from ATCA observations of the radio galaxy PKSB1733-56.
The grey scales are saturated at $\pm800 \mu$Jy beam$^{-1}$ ($\pm0.15$\%
of the peak intensity of the core). The left image used the 
multi-frequency deconvolution algorithm of
{\it Sault and Wieringa} [{\it 1994}], whereas the right image used a standard CLEAN. Because spectral effects of the core
were eliminated by a calibration technique, the right image is limited by the spectral effects of the
{\em second} brightest feature -- the north-east hot spot. Correlator-based errors are believed to limit
the left image. Reproduced from {\it Sault and Wieringa} [{\it 1994}].}
\label{fig:mfs}
\end{figure}

{\it Conway et al.} [{\it 1990}] and {\it Conway and Sault} [{\it 1995}] considered the errors
that result from truncating the spectral index expansion, and
showed that for
a spread of frequencies of $\pm 12.5\%$ and typical spectral indices, the
resultant errors
are of order $5\times10^{-4}$ times the {\em second} brightest feature in the
image. This assumes that the spectral variation of the brightest source
has been eliminated by calibration or self-calibration of the
spectral flux scale. {\it Sault and Wieringa} [{\it 1994}] considered effects, other than source
spectral index, which cause spectral variation, and gave details of a practical
multi-frequency synthesis experiment, including some self-calibration
issues. Their experiment observed multiple frequencies
in three ways: the input bandwidth was broken into 32 correlator channels,
two frequencies were observed simultaneously, and these two frequencies
were altered every 75 s. By using multi-frequency synthesis,
Sault and Wieringa were able to obtain good $u-v$ coverage from
two configurations with limited antennas.

\section{The Short Spacing Problem and Mosaicing}\label{sec:mosaic}
One of the blessings, and one of the curses, of interferometry is its
insensitivity to very extended features. This is readily seen: an
interferometer array has a minimum spacing, and structure
that is broad enough to produce insignificant flux at the shortest
spacing cannot be readily detected. In this regard an interferometer
array acts as a filter to broad structure.
The antenna dish diameter, $D$, places a limit on the minimum spacing,
as shorter spacings would result either in shadowing or, worse still, in
physical
collision of antennas. In practice, the minimum spacing, $d_{\rm min}$, for 
most arrays is somewhat larger than $D$.

Although deconvolution algorithms can go some way towards estimating short
spacings, these are usually inadequate when the extended emission is complex
and significant. Measuring the short spacings directly is required.
This is commonly done with a single-dish telescope the diameter of which is
at least twice
the minimum interferometer spacing (a single-dish telescope
is sensitive to spacings out to its diameter). A larger diameter than
the minimum spacing is required, both because single-dish illumination patterns
are typically heavily tapered (and so there is poor sensitivity at the spacings
corresponding to the single-dish diameter) and to provide an overlap 
region to allow simple gain
cross-calibration between the interferometer and single-dish data.
Two approaches have traditionally been used to combine the interferometer
and single-dish data. The first is to
form short spacing, pseudo-interferometer visibilities
from the single-dish data. These
visibilities are then used in the normal interferometric imaging procedures.
Because of the different
imaging equations of single-dish and interferometer images,
some manipulations are required to form the pseudo-interferometer
visibilities. See {\it Bajaja and van Albada} [{\it 1979}] for details. The second
traditional approach -- the so-called feathering approach -- combines
images. The assumption is that final, deconvolved interferometer and single-dish
images have been formed. Each image provides an accurate representation of
the true sky in only a limited region of the $u-v$ plane. The true
sky can be formed by combining the Fourier transforms
of the two images, using the short spacings in the single-dish
image and the long spacings of the interferometric image
[e.g. {\it Higgs et al.}, {\it 1977}].
The overlap region in the $u-v$ can again be used for simple gain
cross-calibration, and can be averaged in some way in forming the output.
{\it Holdaway} [{\it 1992a}] reviewed this approach in a somewhat
different context. {\it Schwarz and Wakker} [{\it 1991}] discussed a variant, as well
as comparing the two traditional approaches.

Another characteristic of interferometer arrays
is that their field of view is limited by the primary beam response.
Sources larger than the primary beam size
must be imaged by using multiple pointings of the interferometer array, and
the different fields pieced together.
Note that the primary
beam extent and the short-spacing problem are both set by $D/\lambda$ -- both
effects indicate that it is not straightforward to image structures larger
than this size.

Only in the last decade has it been recognized that both
these shortcomings of an interferometer can be overcome (within limits) by
`mosaicing' -- the practice of forming a single image from multiple pointings
of an interferometer [{\it Cornwell}, {\it 1989} gives an overview].
Although mosaicing is clearly needed to image
sources larger than the primary beam, it is also effective in measuring
short spacings. This stems from the fact that a single
interferometer with baseline $d$ does not measure just a single spacing
in the $u-v$ plane -- it measures an integral of
spacings from $d-D$ to $d+D$. This means, in principle, that if the projected
spacing between two antennas is the minimum ($d_{\rm min} \approx D$),
then that interferometer is sensitive down to
(but excluding) the zero spacing.

Whereas the information
in this range of spacings is generally not accessible in a single pointing,
it can be recovered in a multi-pointing observation. This was first
noted by {\it Ekers and Rots} [{\it 1979}], and the argument substantially enhanced
by {\it Cornwell} [{\it 1988}]. Using Cornwell's argument, if the pointing centre,
$(\ell_{\rm P},m_{\rm P})$, is treated as a variable in the imaging
equation, then the measured visibilities are given by
\begin{equation}
V(u,v;\ell_{\rm P},m_{\rm P}) = \int A(\ell-\ell_{\rm P},m-m_{\rm P})I(\ell,m)\exp(-i2\pi(u\ell+vm))d\ell dm.
\end{equation}
Here, for simplicity, we assume that the delay-tracking
centre remains fixed at the origin. Let $a(u,v)$ and $i(u,v)$ be the
Fourier transforms of the primary beam pattern and the true sky brightness, then
it is straightforward to show that
\begin{equation}
\int V(u,v;\ell_{\rm P},m_{\rm P})\exp(-i2\pi(\ell_{\rm P}\Delta u+m_{\rm P}\Delta v )d\ell_{\rm P}dm_{\rm P} = a(\Delta u,\Delta v)i(u+\Delta u,v+\Delta v).\label{eq:pntfft}
\end{equation}
This can be seen as a Fourier transform of $V$ from the 
$(\ell_{\rm P},m_{\rm P})$ domain to the $(\Delta u,\Delta v)$ domain.
So, in principle, by measuring the visibility over a number of pointings,
$i(u+\Delta u,v+\Delta v)$ can be recovered for $(\Delta u,\Delta v)$ 
out to $D/\lambda$.

{\it Cornwell} [{\it 1988}] showed that scanning was not required -- all
information can be recovered if the pointings are sampled on a grid, the
increment of which is
no larger than $\lambda/2D$ radians (this is typically near the half-power
point of the primary beam response).
This results from a Nyquist
sampling criterion, which applies
because the antenna diameter is finite and so the 
response of an interferometer is inherently
band-limited in the spatial frequency domain.

{\it Cornwell} [{\it 1988}] presented a practical algorithm, based on MEM,
which does a joint deconvolution of all the separate pointings. Basically the algorithm searches for
a sky brightness distribution which maximizes an entropy measure, but which
is compatible with the dirty images formed at each pointing.
The algorithm is similar to that in
Section 3.2
except that both the primary beam
and the synthesized beam must be accounted for when converting from a model of the
sky to a dirty image at one pointing. The recovery of the extended emission is
implicit in the algorithm -- Fourier transforms such as
Equation (14)
never explicitly appear. To date, observations containing
several hundred pointings have been deconvolved using Cornwell's approach
(e.g. {\it Staveley-Smith et al.} [{\it 1995}], see
Figure 3).

Although, in principle, mosaicing should recover spacings of $\pm D/\lambda$
around each
sample in the $u-v$ plane, Cornwell reported that, in practice, the
improvement is closer to half this.
In part, this undoubtedly results from the reduced sensitivity as
$a(\Delta u, \Delta v)$ drops to zero at $\pm D/\lambda$. So, to
recover all spacings effectively, single-dish data are still required. Cornwell's
algorithm also incorporates single-dish measurements in an elegant fashion.
They are simply extra data (albeit with a different imaging equation) that
the solution needs to be consistent with.

Since the {\it Cornwell} [{\it 1988}] paper, extensive research and
simulation work on mosaicing has been done. This is particularly so within the
NRAO, where
the driving force has come from planning for the proposed NRAO millimeter
array (mosaicing is important at millimeter wavelengths, because the
emission processes produce extended structures and because of smaller
primary beam sizes).
The NRAO Millimeter Array Memorandum Series contains many papers
on mosaic-related issues. A summary of a number of aspects of the
work is presented in {\it Cornwell et al.} [{\it 1993}]. Three imaging-related
results to come out of this research will be briefly described. 

\begin{figure}
\begin{center}\epsfysize=9cm\leavevmode\epsffile{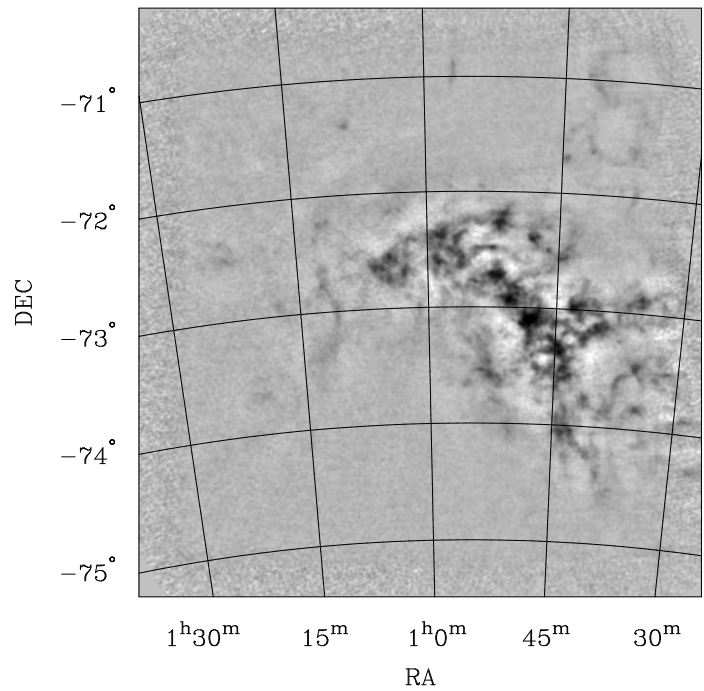}\end{center}
\caption[]{
An HI channel image of a mosaic of the Small Magellanic Cloud.
This observation consisted of 320 pointings, with 22 m antennas (primary
beam size of $34\arcmin$), and minimum spacing of 31 m. The resolution is
$100\arcsec$. The largest residual errors in this image result
from no single-dish data being used. Reproduced from {\it Staveley-Smith et al.} [{\it 1995}].}
\label{fig:mosaic}
\end{figure}

\subsection{The homogeneous array concept}
One aim of the millimeter array design is to measure {\em all} spacings
up to some maximum.
Given that mosaicing interferometer data cannot produce good estimates
for spacings
shorter than about $D/2$, there has been considerable debate about the best method to
measure these. Although {Cornwell et al.} [{\it 1993}]
mention several possibilities, only two are serious contenders:
the large single-dish approach (with its diameter a few times
the minimum interferometer spacing), and the so-called homogeneous array
(with each element of the interferometer acting as a single dish, and with the
minimum interferometer spacing slightly larger than $D$). With the
homogeneous array, it is assumed that the spacings down to about $D/2$
are determined interferometrically using joint deconvolution,
whereas the antennas in single-dish
mode provide information on spacings shorter than this.

{\it Cornwell et al.} [{\it 1993}] concluded that both the homogeneous array and large
single-dish approaches can produce high-quality results. In particular, they
showed that the homogeneous array is reasonably robust to a number of
instrumental errors. Given the greater simplicity of the homogeneous array,
and the greater ease of achieving the pointing and surface accuracy
requirements with small elements, the homogeneous array approach is to be
preferred. 

\subsection{Errors in mosaicing experiments}
In interferometric, single-pointing images of objects smaller than the
primary beam, ``primary beam'' errors (pointing errors, errors in the assumed
form of the primary beam, errors caused by a primary
beam which is not circularly symmetric, etc) are of limited importance.
Indeed the final dynamic range of images is
totally independent of some primary beam errors (e.g. a constant
pointing offset in RA and DEC). This is not so
with mosaicing -- a good model of the overall primary beam response
is as important as a good model of the synthesized beam. Both the
primary and synthesized beams are important in determining the overall
PSF of a mosaic experiment.

{\it Cornwell et al.} [{\it 1993}] presented formulae which estimate the effects of
various errors in the resultant final images. They also presented
simulation results to confirm the validity of the formulae.
For example, they suggested
that if $\sigma_{\rm B}$ is the rms sidelobe level in the synthesized beam,
and if $\sigma_{\rm A}$ is the rms difference between the actual and 
assumed primary beam response,
and if the primary beam responses of all antennas are identical, then the
dynamic range in an image will be limited to approximately
\begin{equation}
\frac{1}{\sigma_{\rm A}\sigma_{\rm B}}.
\end{equation}
This indicates the importance, both of knowing the primary beam well,
and for an array which has good $u-v$ coverage to minimize the synthesized
beam sidelobes.

Some research has investigated ways to reduce the effects of pointing
errors in the final images. {\it Holdaway} [{\it 1993}] presented an
algorithm which can handle time-varying, but known, pointing errors in
interferometer data. {\it Emerson} [{\it 1991}]
demonstrated a cross-calibration procedure for determining 
global pointing errors in single-dish data, which he showed is a
significant error
when combining interferometer and single-dish data.
Emerson's results show that, left
uncorrected, global single-dish pointing errors are more significant for
larger dishes. However, they are easier to correct in this case, and
the resultant performance is superior to that for a smaller dish. His
results suggest that there was some advantage in using the large
single-dish approach
rather than the homogeneous array.
Presumably this is because there is a greater region of overlap
in the $u-v$ domain in which to cross-calibrate for the former than for
the latter.

Other references to primary beam errors include {\it Braun} [{\it 1988b}]
and {\it Holdaway} [{\it 1992b}].

\subsection{Alternative imaging approaches}
{\it Cornwell} [{\it 1994}] and {\it Cornwell et al.} [{\it 1993}] provide reviews of alternative
mosaicing algorithms. All mosaicing schemes, at some level, perform a form
of linear
mosaicing operation (a weighted sum of the pixels in the images of the
different pointings). If $I_p(\ell,m)$ is an image of the
$p$th pointing, then the linear mosaic is
\begin{equation}
I_{\rm LM}(\ell,m) =
\frac{\sum_p A(\ell-\ell_p,m-m_p)I_p(\ell,m)/\sigma^2_p}{\sum_p A^2(\ell-\ell_p,m-m_p)/\sigma^2_p}.
\end{equation}
Here $\sigma^2_p$ is the noise level in image of each pointing.

One approach to mosaicing is to deconvolve images of each pointing
individually, and then to mosaic the results linearly. The greater
computational simplicity of this approach makes it attractive for
large-sky surveys [e.g. {\it Condon et al.}, {\it 1994}; {\it Bremer}, {\it 1994}]. Primary
beam errors also have less effect on the dynamic range of such images, as
the deconvolution is done in single-pointing mode. However,
the short-spacing advantages of mosaicing can only be realized in
joint deconvolution (i.e. if the
multiple pointings are combined before or during the deconvolution
process) [{\it Cornwell}, {\it 1988}].

{\it Sault et al.} [{\it 1995}] and {\it Viallefond and Guilloteau} [{\it 1993}] take
a different approach. Both form a linear mosaic of the dirty
images of each pointing, and then deconvolve this composite
image. They do this in a way which accounts for the position-variant
nature of the PSF. Sault et al. show
that there are practical advantages to such an approach and that it
is capable of recovering short-spacing information. Buit whereas Sault et al.
present deconvolvers based on the maximum entropy
method and the {\it Steer et al.} [{\it 1984}] variant of CLEAN,
Viallefond and Guilloteau use a more traditional {\it H\"ogbom} [{\it 1974}] CLEAN algorithm.

{\it Braun} [{\it 1988a}, {\it 1988b}] noted that, as dynamic range in joint deconvolution
can be limited by primary beam errors on the bright compact objects, and
as these objects benefit little from joint deconvolution, then a hybrid
imaging scheme may be desirable. He advocated using a joint deconvolution
approach with the shorter interferometer spacings and single-dish data,
and individual imaging and deconvolution for the longer spacing
data. The two resultant images are then combined using a traditional feathering
operation. In addition to dynamic range considerations, this scheme may be
computationally cheaper. {\it Holdaway} [{\it 1992a}] expanded on some of Braun's
approaches.

\section{Wide-field imaging with non-coplanar baselines}\label{s:noncoplanar}
In the two-dimensional Fourier transform approximation of
Equation (2),
the $w(n-1)$ term of the van Cittert-Zernike equation has been
neglected. There are two regimes where this is appropriate.
\begin{enumerate}
\item If the sampling in $u-v-w$ is coplanar, i.e.
\begin{equation}
w=au+bv
\end{equation}
for some $a$ and $b$,
then the van Cittert-Zernike equation reduces
to a two-dimensional Fourier relation in the distorted `direction cosines'
$\ell',m'$
\begin{eqnarray}
\ell' &=& \ell + a\left(\sqrt{1-\ell^2-m^2} - 1\right),\\
 m' &=& m + b\left(\sqrt{1-\ell^2-m^2} - 1\right).
\end{eqnarray}
In practice these distorted direction cosines mean simply that a
different set of formulae are required
to convert from pixel coordinates to celestial coordinates.
Coplanar observations will result from east-west arrays
and from snapshots of physically planar arrays.
\item
For observations with non-coplanar baselines, the two-dimensional approximation
is equivalent to ignoring the phase term,
\begin{equation}
\phi = 2\pi (n-1) w.
\end{equation}
For small fields ($n\approx1$) this may be adequate.
\end{enumerate}
The first regime involves no approximation; the techniques of this section
are not relevant for east-west arrays.
We are interested here in eliminating the approximation in the second regime.

The phase term ignored in the two-dimensional approximation is
\begin{eqnarray}
\phi &=& 2\pi (n-1)w\\
     &\approx& -\pi (\ell^2+m^2)w.
\end{eqnarray}
Assuming that a two-dimensional Fourier transform is used, the effect of this
approximation is to smear sources in proportion to the square of their
distance from the delay-tracking centre. The smearing, being
a property of array geometry etc only, is predictable for a given observation;
it can be viewed as a position-variant PSF when the conventional
approach is used to transform. Near the centre, the conventional
dirty beam represents the PSF well, whereas at
a distance from the centre, its representation is poor.

The limitations of ignoring the phase term have been appreciated
for some time and algorithms have been proposed to account for it
[e.g. {\it Brouw}, {\it 1975};
{\it Clark}, {\it 1973}, {\it 1978}; {\it Frater and Docherty}, {\it 1980};
{\it Bracewell}, {\it 1984}; {\it McLean}, {\it 1984}]. However, it was not until
recently that astrophysical requirements and improved computational
resources have led to the implementation of these algorithms. This is,
in part, a result of the development of high-resolution, low-frequency
arrays. To see that the problem is more significant at these frequencies,
{\it Cornwell and Perley} [{\it 1992}] noted that the maximum phase term goes as
\begin{equation}
\phi \sim \frac{\lambda d_{\rm max}}{D^2}.
\end{equation}
This assumes that emission fills the primary beam.
As this is often not the case at shorter wavelengths ($\lambda < 20$ cm),
the importance of the effect is steeper than the linearity that the above
formula suggests.

Algorithms to handle the problem can be divided into four broad classes: we
will discuss each in turn. {\it Cornwell and Perley} [{\it 1992}] gave an overview of the
problem, and compared some approaches. They also discussed the implementation
of two approaches in detail.
Examples from this paper are given in
Figure 4.

\begin{figure}
\begin{center}
\epsfysize=4.5cm\leavevmode\epsffile{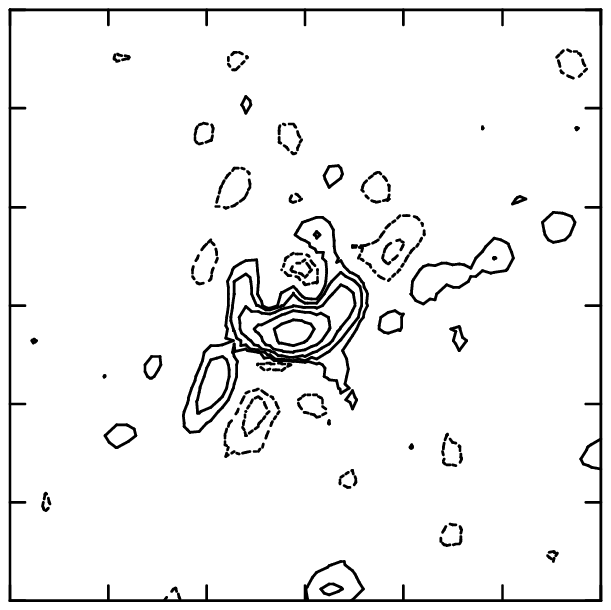}\hskip 5mm
\epsfysize=4.5cm\leavevmode\epsffile{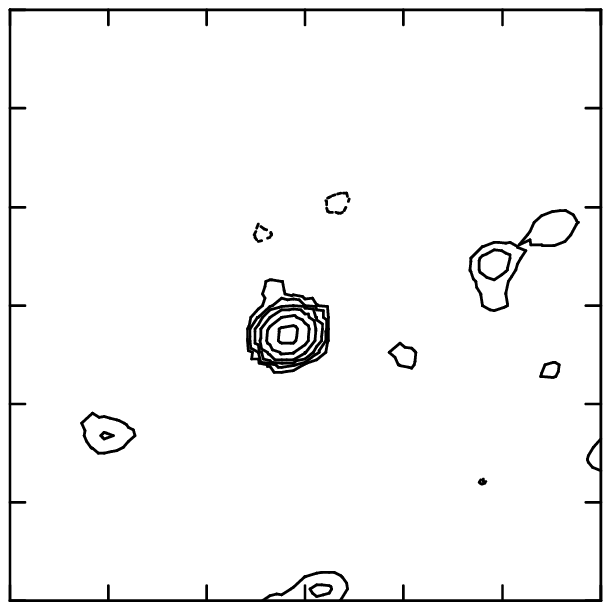}\hskip 5mm
\epsfysize=4.5cm\leavevmode\epsffile{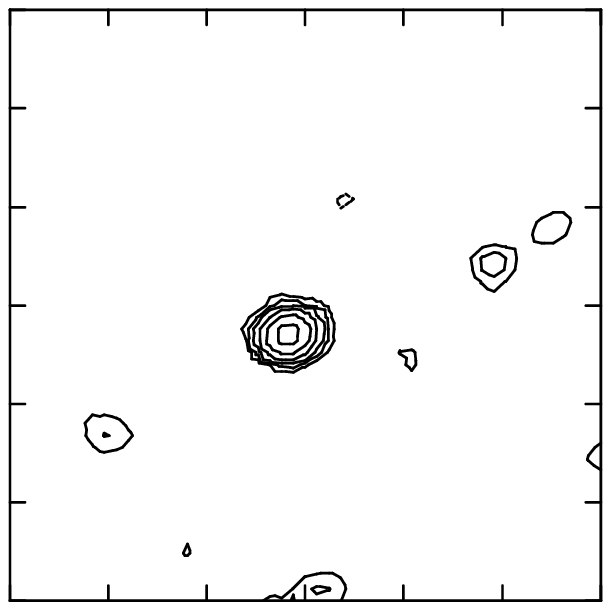}
\end{center}
\caption[]{
These images show the effects of non-coplanar baselines. They show a deconvolved background object
about $2\degr$ from the tangent point of a VLA observation at
327 MHz: (left) conventional two-dimensional approach;
(centre) three-dimensional transform method; (right) polyhedron method. The resolution is 
$160\arcsec\times160\arcsec$. The tick marks are spaced by $500\arcsec$. Contours levels are
1, 2, 4, 8 and 16 times $\pm15$ mJy beam$^{-1}$. Reproduced from
{\it Cornwell and Perley} [{\it 1992}].}
\label{fig:noncoplanar}
\end{figure}

\subsection{Deconvolution with a position-variant PSF}
This approach was originally proposed by {\it McLean} [{\it 1984}], whereby a CLEAN-like
algorithm is used to deconvolve a smeared image: when a point source is
located in a CLEAN iteration, the true PSF at that position is computed and
subtracted. McLean proposed a power-series expansion of dirty beams
to determine the 
PSF. {\it Waldram and McGilchrist} [{\it 1990}] and {\it Waldram} [{\it 1991}]
implemented a conceptually similar
approach, although they computed the position-variant PSFs on a coarse
grid on the sky, and used a weighted sum of these
to approximate the PSF at any point in the field. Their application
was for a linear array (the Cambridge Low Frequency Synthesis Telescope)
which deviates from east-west by only $3\degr$.

\subsection{Approaches using coplanar decomposition}
If visibility sets that are coplanar do not have a problem, one
approach is to decompose a non-coplanar set of visibilities into
a collection of (at least approximately) coplanar sets. Each coplanar
set could then be imaged, deconvolved, and distorted onto a common image
grid. This was the basis of suggestions by {\it Bracewell} [{\it 1984}] and 
{\it Frater and Docherty} [{\it 1980}], although their decompositions, motivated
by different array geometries, were quite different.
For a physically coplanar array, decomposing into a set of snapshots is
another possibility. This was the approach taken by {\it Condon et al.} [{\it 1994}] in
a somewhat different context. The main shortcoming with this approach
is that, as deconvolution is usually a non-linear operation, the sum
of the separate deconvolutions will be inferior to a single `correct'
deconvolution.

\subsection{The polyhedron method}
If the phase term can be ignored for sufficiently small fields, one
approach is to build up a large field from small facets of the
sky -- i.e. the celestial sphere is approximated by a celestial
polyhedron. This approach -- the so-called polyhedron method -- appears to have been first suggested by
{\it Clark} [{\it 1979}], and has been implemented and discussed in detail by
{\it Cornwell and Perley} [{\it 1992}]. The Cornwell and Perley implementation, in many
respects, is an extension of the Cotton-Schwab CLEAN algorithm [{\it Schwab}, {\it 1984}].

\subsection{Three-dimensional approach}
This approach concedes that the problem is not fundamentally two
dimensional. Instead it notes that simple manipulation can convert
the van Cittert-Zernike equation to a three-dimensional Fourier
transform
\begin{equation}
V(u,v,w)\exp(-i2\pi w) =
\int \frac{I(\ell,m)\delta(\sqrt{1-\ell^2-m^2}-n)}{\sqrt{1-\ell^2-m^2}}
\exp\left(-i2\pi(u\ell+vm+wn)\right) d\ell\,dm\,dn.
\end{equation}
Here the visibility (uncorrected for delay tracking) is a three-dimensional
transform of the celestial sphere.
This is probably the most intuitive and first-recognized approach
[e.g. {\it Clark}, {\it 1973}; {\it Brouw}, {\it 1975}]. {\it Perley} [{\it 1989}] and {\it Cornwell and Perley} [{\it 1992}]
considered the technique and its implementation.
They showed that a three-dimensional dirty
beam relates the true and dirty celestial spheres by a convolution
relationship. They generalized two-dimensional deconvolution algorithms to
work on this three-dimensional problem. In practical terms
{\it Cornwell and Perley} [{\it 1992}] noted that, whereas a three-dimensional transform
approach is computationally cheap and simpler to program compared to the
polyhedron method, it is expensive in terms of memory.

\section{Non-isoplanatic imaging}
In all the preceding discussion, we implicitly assumed
that the visibility data were corrected for antenna gains. However,
to be able to do this
requires isoplanaticism -- the assumption that the atmospheric
path delays are constant over the field of view. That is,
the antenna gains are independent of source position. This, however,
will not
be the case at wavelengths longer than about 1 m, because of large
ionospheric irregularities and large primary beam sizes. Non-isoplanaticity
will be significant, for example, for the GMRT [{\it Swarup}, {\it 1990}, {\it 1991}]
and for the VLA at 74 MHz [{\it Kassim et al.}, {\it 1993}].

Non-isoplanaticism will mean that standard imaging, calibration and
self-calibration schemes will be inadequate. {\it Schwab} [{\it 1984}] considered
extensions to self-calibration, where an antenna phase was determined
at a number of points on the sky, and interpolation used for intermediate
positions. He also advocated a modified Cotton-Schwab CLEAN algorithm similar
to the polyhedron method of {\it Cornwell and Perley}
[{\it 1992} -- see
Section 6
above] for imaging and deconvolution.

{\it Subrahmanya} [{\it 1990}, {\it 1991}] advocated a different self-calibration model:
because the primary beams of the different antennas of the VLA and GMRT
overlap significantly at typical ionospheric heights, he suggested that
the number of unknowns in a self-calibration solution could be reduced
by solving for a geometric model of the ionosphere. The ionosphere would be
modelled as a number of phase cells at a given height. The effect of
these cells on the observed phase of a given source on a given
antenna could then be predicted. {\it Bhatnagar} [{\it 1995}] reports that convergence
of Subrahmanya's method is very sensitive to the ionospheric cell size
and the initial model of the source.

Although {\it Kassim et al.} [{\it 1993}] reported VLA observations at 74 MHz, they
restricted their attention to strong, relatively compact, sources where
non-isoplanatic issues could be ignored to first order. As yet, there
appear to be no practical implementations of non-isoplanatic imaging
algorithms.

\section{Acknowledgements}
We thank the referees, W. N. Brouw, T. J. Cornwell and J. E. Noordam, as well
as R. D. Ekers and D. Goddard for their comments and corrections.
We are also grateful to D. S. Briggs, T. J. Cornwell
and L. Staveley-Smith for allowing us to reproduce diagrams from their papers.

\section{References}
\setlength{\parindent}{0pt}
\setlength{\parskip}{2.5mm}
J. G. Ables [1974], ``Maximum entropy spectral analysis,''
{\it Astr. Astrophys. Suppl.}, {\bf 15}, pp. 383-393.

H. C. Andrews, and B. R. Hunt [1977], {\it Digital Image Restoration, Signal Processing},
Englewood Cliffs, New Jersey, Prentice-Hall Inc.

E. Bajaja, and G. D. van Albada [1979], ``Complementing aperture synthesis radio
data by short spacing components from single dish observations,''
{\it Astr. Astrophys.}, {\bf 75}, pp. 251-254.

S. Bhatnagar [1995], GMRT Internal Report, Pune, India.

R. N. Bracewell [1984], ``Inversion of nonplanar visibilities,''
in J. A. Roberts (ed.), {\it Proceedings of IAU/USRI Conference on
Indirect Imaging}, Cambridge Univ. Press, pp. 177-183.

R. Braun [1988a], ``Mosaicing with high dynamic range,''
NRAO Millimeter Array Memo 46, Socorro, NM.

R. Braun [1988b], ``Simulations of primary beam truncation and pointing errors
with the MMA,'' NRAO Millimeter Array Memo 54, Socorro, NM.

M. A. R. Bremer
[1994], ``The Westerbork Northern Sky Survey (WENSS): A radio survey using
the mosaicing technique,'' in D. R. Crabtree, R. J. Hanisch, and J. Barnes (eds.),
{\it Astronomical Data Analysis Software and Systems III}, Astronomical Society of the Pacific
Conference Series, {\bf 61}, pp. 175-178.

A. H. Bridle, and F. R. Schwab [1989], ``Wide Field Imaging I: Bandwidth and
time-averaging smearing,''
in R. A. Perley, F. R. Schwab, and A.H. Bridle (eds.),
{\it Synthesis Imaging in Radio Astronomy}, Astronomical Society of the Pacific
Conference Series, {\bf 6}, pp. 247-258.

Briggs D.S. [1995], {\it Ph.D. Thesis}, New Mexico Institute of Mining and
Technology.

E. Brinks, and W. W. Shane [1984], ``A high resolution hydrogen line
survey of Messier 31. I. Observations and data reduction,''
{\it Astr. Astrophys. Suppl.}, {\bf 55}, pp. 179-251.

W. N. Brouw [1975], ``Aperture synthesis,''
in B. Alder, S. Fernbach, M. Rotenberg (eds.),
{\it Methods in Computational Physics}, {\bf 14}, pp. 131-175.

B. G. Clark [1973], ``Curvature of the sky,''
VLA Scientific Memo 107, NRAO, Socorro, NM.

B. G. Clark [1975], ``Bandwidth correction by deconvolution,''
VLA Scientific Memo 118, NRAO, Socorro, NM.

B. G. Clark [1979], ``Digital processing methods for aperture synthesis
observations,'' in C. van Schooneveld (ed.), {\it Proceedings of IAU
Colloquium 49, Image Formation from Coherence Functions in Astronomy}, D.Reidel, pp. 113-120.

B. G. Clark [1980], ``An efficient implementation of the algorithm CLEAN,''
{\it Astr. Astrophys.}, {\bf 89}, pp. 377-378.

J. J. Condon, W. D. Cotton, E. W. Greisen, Q. F. Yin, R. A. Perley,
and J. J. Broderick [1994], ``The NRAO VLA Sky Survey,''
in D. R. Crabtree, R. J. Hanisch, and J. Barnes (eds.), {\it Astronomical
Data Analysis Software and Systems III}, Astronomical Society of the Pacific
Conference Series, {\bf 61}, pp. 155-164.

J. E. Conway [1991], ``Multi-frequency synthesis,'' in T. J. Cornwell, and
R. A. Perley (eds.),
{\it Proceedings of IAU Colloquium 131, Radio Interferometry: Theory, Techniques and Applications}, Astronomical Society of the Pacific Conference Series,
{\bf 19}, pp. 171-179.

J. E. Conway, and R. J. Sault [1995], ``Multi-frequency synthesis,''
in J. A. Zensus, P. J. Diamond, and P. J. Napier (eds.),
{\it Workshop on Very Long Baseline Interferometry and the VLBA},
Astronomical Society of the Pacific Conference Series, {\bf 82}, pp. 309-325.

J. E. Conway, T. J. Cornwell, and P. N. Wilkinson [1990], ``Multi-frequency
synthesis: a new
technique in radio interferometric imaging,'' {\it Mon. Not. R. Astr. Soc.}, {\bf 246}, pp. 490-509.

T. J. Cornwell [1983], ``A method of stabilizing the clean algorithm,''
{\it Astr. Astrophys.}, {\bf 121}, pp. 281-285.

T. J. Cornwell [1984], ``Broadband mapping of sources with spatially
varying spectral index,'' VLB Array Memo 324, NRAO, Socorro, NM.

T. J. Cornwell [1987], ``Radio-interferometric imaging of weak objects in
conditions of poor phase stability: the relationship between speckle masking
and phase closure methods,'' {\it Astr. Astrophys.}, {\bf 180}, pp. 269-274.

T. J. Cornwell [1988], ``Radio-interferometric imaging of very large
objects,'' {\it Astr. Astrophys.}, {\bf 202}, pp. 316-321.

T. J. Cornwell [1989], ``Wide field imaging III: mosaicing,''
in R. A. Perley, F. R. Schwab, and A.H. Bridle (eds.),
{\it Synthesis Imaging in Radio Astronomy}, Astronomical Society of the Pacific
Conference Series, {\bf 6}, pp. 277-286.

T. J. Cornwell [1994], ``Mosaicing: Current status,'' in M. Ishiguro, 
and W. J. Welch (eds.), {\it Astronomy with Millimeter
and Submillimeter Wave Interferometry}, Astronomical Society of the Pacific Conference Series,
{\bf 59}, pp. 96-103.

T. J. Cornwell, and R. Braun [1989], ``Deconvolution,''
in R. A. Perley, F. R. Schwab, and A. H. Bridle (eds.),
{\it Synthesis Imaging in Radio Astronomy}, Astronomical Society of the Pacific Conference Series,
{\bf 6}, pp. 167-184.

T. J. Cornwell, and E. B. Fomalont [1989], ``Self-calibration,''
in R. A. Perley, F. R. Schwab, and A. H. Bridle (eds.),
{\it Synthesis Imaging in Radio Astronomy}, Astronomical Society of the Pacific Conference Series,
{\bf 6}, pp. 185-197.

T. J. Cornwell, M. A. Holdaway, and J. M. Uson [1993], ``Radio-interferometric
imaging of very large objects: implications for array design,''
{\it Astr. Astrophys.}, {\bf 271}, pp. 697-713.

T. J. Cornwell, and R. A. Perley [1992], ``Radio-interferometric imaging of very
large fields -- The problem of non-coplanar arrays,'' {\it Astr. Astrophys.}, 
{\bf 261}, pp. 353-364.

T. J. Cornwell, and P. N. Wilkinson [1981], ``A new method for making maps with
unstable radio interferometers,'' {\it Mon. Not. R. Astr. Soc.}, {\bf 196}, pp. 1067-1086.

W. D. Cotton [1989], ``Special problems in imaging,''
in R. A. Perley, F. R. Schwab, and A. H. Bridle (eds.),
{\it Synthesis Imaging in Radio Astronomy}, Astronomical Society of the Pacific Conference Series,
{\bf 6}, pp. 233-246.

R. D. Ekers [1983], ``The almost serendipitous discovery of self-calibration,''
in K. Kellermann, and B. Sheets (eds.),
{\it Proceedings of the Workshop on Serendipitous Discoveries in
Radio Astronomy}, NRAO.

R. D. Ekers, and A. H. Rots [1979], ``Short spacing synthesis from a primary beam
scanning interferometer,'' in C. van Schooneveld (ed.), {\it Proceedings of
IAU Colloquium 49, Image Formation from Coherence Functions in Astronomy}, D. Reidel, pp. 61-63.

D. T. Emerson [1991], ``Imaging characteristics of the proposed NRAO MMA,
with and without a single large element, and a pointing correction
algorithm,'' in T. J. Cornwell, and R. A. Perley (eds.),
{\it Proceedings of IAU Colloquium 131, Radio Interferometry: Theory, Techniques and Applications}, Astronomical Society of the Pacific
Conference Series, {\bf 19}, pp. 441-444.

R. H. Frater, and I. S. Docherty [1980], ``On the reduction of three
dimensional interferometer measurements,'' {\it Astr. Astrophys.}, {\bf 84}, pp. 75-77.

P. W. Gorham, A. M. Ghez, S. R. Kulkarni, T. Nakajima., G. Neugebauer,
J. B. Oke, and T. A. Prince [1989], ``Diffraction-limited imaging. III 30 mas
closure phase imaging of six binary stars with the Hale 5m telescope,''
{\it Astron. J.}, {\bf 98}, pp. 1783-1799.

S. F. Gull, and G. J. Daniell [1978], ``Image reconstruction from incomplete
and noisy data,'' {\it Nature}, {\bf 272}, pp. 686-690.

C. A. Haniff, C. D. Mackay, D. J. Titterington, D. Sivia, and J. E. Baldwin [1977],
``The first images from optical aperture synthesis,'' {\it Nature}, {\bf 328}, pp. 694-696.

L. A. Higgs, T. L. Landecker, and R. S. Rogers [1977], ``The true extent of
the Gamma Cygni supernova remnant,''
{\it Astron. J.}, {\bf 82}, pp. 718-724.

J. A. H\"ogbom [1974], ``Aperture synthesis with a non-regular distribution
of interferometer baselines,'' {\it Astr. Astrophys. Suppl.}, {\bf 15}, pp. 417-426.

M. A. Holdaway [1992a], ``Mosaicing with even higher dynamic range,''
NRAO Millimeter Array Memo 73, Socorro NM.

M. A. Holdaway [1992b], ``Surface accuracy requirements for mosaicing
at millimeter wavelengths,'' NRAO Millimeter Array Memo 74, Socorro NM.

M. A. Holdaway [1993], ``Imaging with known pointing errors,''
NRAO Millimeter Array Memo 95, Socorro NM.

R. C. Jennison [1951], {\it Ph.D. Thesis}, University of Manchester.

R. C. Jennison [1958], ``A phase sensitive interferometer technique for the
measurement of the Fourier transforms of spatial brightness distributions
of small angular extent,'' {\it Mon. Not. R. Astr. Soc.}, {\bf 118}, pp. 276-284.

N. E. Kassim, R. A. Perley, W. C. Erickson, and K. S. Dwarakanath [1993],
``Subarcminute resolution imaging of radio sources at 74 MHz with
the Very Large Array,'' {\it Astron. J.}, {\bf 106}, pp. 2218-2228.

A. Lannes, E. Anterrieu, and K. Bouyoucef [1994], ``Fourier interpolation and
reconstruction via Shannon-type techniques. I. Regularization principle,''
{\it J. Mod. Opt.}, {\bf 41}, pp. 1537-1574.

C. L. Lawson, and R. J. Hanson [1974], {\it Solving Least Squares Problems},
Englewood Cliffs, New Jersey, Prentice-Hall Inc.

A. W. Lohmann, and B. Wirnitzer [1984], ``Triple correlations,'' {\it Proc. IEEE},
{\bf 72}, pp. 889-901.

A. W. Lohmann, G. Weigelt, and B. Wirnitzer [1983], ``Speckle masking in
astronomy: triple correlation theory and applications,'' 
{\it Applied Optics}, {\bf 22}, pp. 4028-4037.

L. L. McCready, J. L. Pawsey, and R. Payne-Scott [1947], ``Solar radiation
at radio frequencies and its relation to sunspots,'' {\it Proc. Roy. Soc. A},
{\bf 190}, pp. 357-375.

D. J. McLean [1984], ``A simple expansion method for wide-field mapping,''
in J. A. Roberts (ed.), {\it Proceedings of IAU/USRI Conference on
Indirect Imaging}, Cambridge Univ. Press, pp. 185-191.

T. Nakajima, S. R. Kulkarni, P. W. Gorham, A. M. Ghez, G. Neugebauer,
J. B. Oke, T. A. Prince, and A. C. S. Readhead [1989], ``Diffraction limited imaging. II. Optical
aperture-synthesis imaging of two binary stars,'' {\it Astron. J.}, {\bf 97}, pp. 1510-1521.

R. Narayan, and R. Nityananda [1986], ``Maximum entropy image restoration
in astronomy,'' {\it Ann. Rev. Astr. Ap.}, {\bf 24}, pp. 127-170.

J. E. Noordam, and A. G. de Bruyn [1982], ``High dynamic range mapping of strong
radio sources, with application to 3C84,'' {\it Nature}, {\bf 299}, pp. 597-600.

T. J. Pearson, and A. C. S. Readhead [1984], ``Image formation by
self-calibration in radio astronomy,'' {\it Ann. Rev. Astr. Ap.}, {\bf 22}, pp. 97-130.

R. A. Perley [1989], ``Wide field imaging II: Imaging with non-coplanar arrays,''
in R. A. Perley, F. R. Schwab, and A. H. Bridle (eds.),
{\it Synthesis Imaging in Radio Astronomy}, Astronomical Society of the Pacific Conference
Series, {\bf 6}, pp. 259-275.

R. A. Perley, F. R. Schwab, and A. H. Bridle (eds.) [1989],
{\it Synthesis Imaging in Radio Astronomy}, Astronomical Society of the
Pacific Conference Series, {\bf 6}.

A. C. S. Readhead, and P. N. Wilkinson [1978], ``The mapping of compact radio
sources from VLBI data,'' {\it Astrophys. J.}, {\bf 223}, pp. 25-36.

A. C. S. Readhead, R. C. Walker, T. J. Pearson, and M. H. Cohen [1980], 
``Mapping
radio sources with uncalibrated visibility data,'' {\it Nature}, {\bf 285}, pp. 137-140.

R. J. Sault [1984], ``The weighting of visibility data,''
VLA Scientific Memo 154, NRAO, Socorro, NM.

R. J. Sault, and M. H. Wieringa [1994], ``Multi-frequency synthesis techniques in
radio interferometric imaging,'' {\it Astr. Astrophys. Suppl.}, {\bf 108}, pp. 585-594.

R. J. Sault, L. Staveley-Smith, and W. N. Brouw [1995], ``An approach to
interferometric mosaicing,'' to be submitted to {\it Astr. Astrophys.}.

F. R. Schwab [1980], ``Adaptive calibration of radio interferometer
data,'' {\it Proc. S.P.I.E.}, {\bf 231}, pp. 18-24.

F. R. Schwab [1984], ``Relaxing the isoplanatism assumption in
self-calibration; applications to low-frequency radio interferometry,''
{\it Astron. J.}, {\bf 89}, pp. 1076-1081.

U. J. Schwarz [1978], ``Mathematical statistical description of the iterative
beam removing technique (Method CLEAN),'' {\it Astr. Astrophys.}, {\bf 65}, pp. 345-356.

U. J. Schwarz [1979], ``The CLEAN method -- use, misuse and variations,''
in C. van Schooneveld (ed.), {\it Proceedings of IAU Colloquium 49, Image Formation
from Coherence Functions in Astronomy}, D.Reidel, pp. 261-275.

U. J. Schwarz [1984], ``The reliability of CLEAN maps and the corrugation
effect,'' in J. A. Roberts (ed.), {\it Proceedings of IAU/USRI Conference on
Indirect Imaging}, Cambridge Univ. Press, pp. 255-260.

U. J. Schwarz, and B. P. Wakker
[1991], ``Adding short spacings to synthesis maps in the sky domain,''
in T. J. Cornwell, and R. A. Perley (eds.), {\it Proceedings of IAU Colloquium 131, Radio
Interferometry: Theory, Techniques and Applications}, Astronomical Society of the Pacific Conference Series,
{\bf 19}, pp. 188-191.

R. A. Sramek, and F. R. Schwab [1989], ``Imaging,'' in R. A. Perley, F. R. Schwab,
and A. H. Bridle (eds.), {\it Synthesis Imaging in Radio Astronomy},
Astronomical Society of the Pacific Conference Series, {\bf 6}, pp. 117-138.

J. L. Starck, A. Bijaoui, B. Lopez, and C. Perrier [1994], ``Image
reconstruction by the wavelet transform applied to aperture synthesis,''
{\it Astr. Astrophys.}, {\bf 283}, pp. 349-360.

L. Staveley-Smith, R. J. Sault, D. McConnell, M. J. Kesteven, D. Hatzidimitriou,
K. C. Freeman, and M. A. Dopita [1995], ``An HI mosaic of the Small Magellanic cloud,''
{\it Publ. Astron. Soc. Aust.}, {\bf 12}, pp. 13-19.

D. G. Steer, P. E. Dewdney, and M. R. Ito [1984], ``Enhancements to the
deconvolution algorithm CLEAN,'' {\it Astr. Astrophys.}, {\bf 137}, pp. 159-165.

C. R. Subrahmanya
[1990], ``Non-isoplanatic ionosphere and low-frequency imaging,''
 in J.E. Baldwin, and Wang Shouguan (eds.), {\it Proceedings of URSI/IAU
Symposium on Radio Astronomical Seeing},
International Academic Publishers, pp. 198-201.

C. R. Subrahmanya [1991], ``Low frequency imaging and the non-isoplanatic
atmosphere,'' in T. J. Cornwell, and R. A. Perley (eds.), {\it Proceedings of
IAU Colloquium 131, Radio Interferometry: Theory, Techniques and Applications},
Astronomical Society of the Pacific Conference Series,
{\bf 19}, pp. 218-222.

G. Swarup [1990], ``Giant metrewave radio telescope (GMRT) -- Scientific
objectives and design aspects,'' {\it Ind. J. Radio \& Space Phys.}, {\bf 19}, pp. 493-505.

G. Swarup [1991], ``Giant Metrewave Radio Telescope (GMRT),''
in T. J. Cornwell, and R. A. Perley (eds.), {\it Proceedings of IAU Colloquium 131, Radio
Interferometry: Theory, Techniques and Applications}, Astronomical Society of the Pacific 
Conference Series, {\bf 19}, pp. 376-380.

A. R. Thompson, and R. N. Bracewell [1974], ``Interpolation and Fourier
transformation of fringe visibilities,'' {\it Astron. J.}, {\bf 79}, pp. 11-24.

A. R. Thompson, J. M. Moran, and G. W. Swenson Jr. [1986], {\it Interferometry and
Synthesis in Radio Astronomy}, New York, New York, John Wiley \& Sons.

F. Viallefond, and S. Guilloteau [1993], ``Mosaic deconvolution,'' in
preparation, Observatoire de Meudon, Paris.

B. P. Wakker, and U. J. Schwarz [1988], ``The Multi-Resolution Clean and its
application to the short-spacing problem in interferometry,''
{\it Astr. Astrophys.}, {\bf 200}, pp. 312-322.

E. M. Waldram [1991], ``The use of beam-sets in the analysis of wide-field
maps from the CLFST,'' in T. J. Cornwell, and R. A. Perley (eds.),
{\it Proceedings of IAU Colloquium 131, Radio
Interferometry: Theory, Techniques and Applications}, Astronomical Society of the Pacific Conference Series,
{\bf 19}, pp. 180-183.

E. M. Waldram, and M. M. McGilchrist [1990], ``Beam-sets -- A new approach to the
problem of wide-field mapping with non-coplanar baselines,''
{\it Mon. Not. R. Astr. Soc.}, {\bf 245}, pp. 532-541.

G. Weigelt [1991], ``Triple-correlation imaging in optical
astronomy,'' in E. Wolf (ed.), {\it Progress in Optics}, Amsterdam, North-Holland,
{\bf XXIX}, pp. 295-319.

G. Weigelt, and B. Wirnitzer [1983], ``Image reconstruction by the speckle-masking
method,'' {\it Opt. Lett.}, {\bf 8}, pp. 389-391.

M. H. Wieringa [1991], ``The use of redundancy in interferometry: A comparison of
redundancy and selfcal,''
in T. J. Cornwell, and R. A. Perley (eds.), {\it Proceedings of IAU
Colloquium 131, Radio Interferometry: Theory, Techniques and Applications},
Astronomical Society of the Pacific Conference Series,
{\bf 19}, pp. 192-196.

M. H. Wieringa [1992], ``An investigation of the telescope based
calibration methods REDUNDANCY and SELF-CAL,'' {\it Experimental Astronomy},
{\bf 2}, pp. 203-225.
\end{document}